\documentclass[aps, prl, twocolumn, superscriptaddress, showpacs]{revtex4}
\usepackage[pdftex]{graphicx}
\usepackage{amsmath,amssymb}
\begin{document}

\title{Bipolaronic blockade effect in quantum dots with negative charging energy}
\author{Tie-Feng Fang}
\affiliation{Center for Interdisciplinary Studies, School of Physics, Lanzhou University, Lanzhou 730000, China}
\author{Shu-Feng Zhang}
\affiliation{Institute of Physics, Chinese Academy of Sciences, Beijing 10080, China}
\author{Chun-Jiang Niu}
\affiliation{Center for Interdisciplinary Studies, School of Physics, Lanzhou University, Lanzhou 730000, China}
\author{Qing-feng Sun}
 \email{sunqf@pku.edu.cn}
 \affiliation{International Center for
Quantum Materials, School of Physics, Peking University, Beijing
100871, China}
 \affiliation{Collaborative Innovation Center of
Quantum Matter, Beijing, China}

\date{\today}

\begin{abstract}
We investigate single-electron transport through quantum dots with
negative charging energy induced by a polaronic energy shift. For
weak dot-lead tunnel couplings, we demonstrate a bipolaronic
blockade effect at low biases which suppresses the oscillating
linear conductance, while the conductance resonances under large
biases are enhanced. Novel conductance plateau develops when the
coupling asymmetry is introduced, with its height and width tuned by
the coupling strength and external magnetic field. It is further
shown that the amplitude ratio of magnetic-split conductance peaks
changes from $3$ to $1$ for increasing coupling asymmetry. Though we
demonstrate all these transport phenomena in the low-order
single-electron tunneling regime, they are already strikingly
different from the usual Coulomb blockade physics and are easy to
observe experimentally.
\end{abstract}

\maketitle

\section{I. introduction}
At low temperatures, various nanostructures weakly coupled to external electrodes exhibit the Coulomb blockade of single-electron (SE) tunneling. This phenomenon has already become a classic hallmark of electronic transport through quantum dots (QDs) fabricated in semiconductor heterostructures \cite{Kastner1992}, nanowires \cite{Yang2005}, carbon nanotubes \cite{Bockrath1997}, and even defined in single molecules or atoms \cite{Zimbovskaya2011}. Generally, SE tunneling occurs when the QD chemical potential, tuned by a gate voltage, is aligned with the transport window opening up between the Fermi energies of the source and drain. Otherwise, the transport is Coulomb blockaded. This gives rise to diamond-shaped blockade regions surrounded by regions of SE tunneling in the differential conductance map as a function of the gate and the source-drain voltage. For vanishing source-drain voltages, the linear conductance features periodic Coulomb peaks as the gate voltage is varied. These peaks are approximatively spaced by the charging energy for adding an electron to the QD. Due to the Coulomb repulsion of electrons, the charging energy is positive, representing a key parameter for the Coulomb blockade physics in the SE tunneling regime \cite{Meir1991}. An interesting question then arises: What will the scenario of SE tunneling be if the charging energy becomes negative?

Indeed, the possibility for inverting the sign of the charging energy has been offered by experiments on single-molecular junctions \cite{SM} and suspended carbon nanotubes \cite{SCNT}, where the QDs are characterized by a Holstein coupling to quantized vibrational degrees of freedom. For sufficiently strong Holstein coupling, the induced polaronic shift can overcome the Coulomb repulsion, resulting in a bipolaronic attraction between electrons and thus a negative effective charging energy for the QD. This possibility has spurred some theories exploring the manifestation of negative charging energy in transport, which mainly concern the charge Kondo effect \cite{CK}, pair tunneling \cite{Koch2006,Hwang2007,Wysokinski2010}, and cotunneling \cite{Hwang2007}. However, these high-order transport behaviors are fragile, requiring harsh experimental conditions. For example, the charge Kondo effect only occurs at the particle-hole symmetric point with an extremely small energy scale. Any deviation from the symmetric point will suppress the Kondo correlations. The pair tunneling is usually subject to the exponential Franck-Condon suppression and is difficult to observe from a broad background of other tunneling contributions. It is hence highly desirable to identify transport phenomena characteristic for negative charging energy in the lower-order SE tunneling regime that is easy to access experimentally.

In this paper, we demonstrate, as a counterpart to the conventional Coulomb blockade physics, a blockade effect of the low-order SE tunneling due to the bipolaronic attraction, which brings about strikingly distinct transport spectroscopy. For weak dot-lead tunnel couplings, while the conductance resonances at large biases are enhanced, the conductance at low biases, especially the oscillating linear conductance, are suppressed. Further asymmetrically tuning the coupling can merge certain enhanced and suppressed conductance peaks into a finite plateau. Its height and width are dependent on the coupling strength and external magnetic field. We also find that the magnetic field splits conductance resonances into two peaks with a novel amplitude ratio of $3$ for symmetric couplings and identical amplitude in the asymmetry limit. These phenomena are explained by the unusual behavior of electron occupation on the dot along with the energy resonance conditions for SE tunneling.

\section{II. model Hamiltonian and formulation}
We start from modeling the QD with one localized orbital \cite{SO} coupled to a vibrational mode as well as to the left ($L$) and right ($R$) leads by the Anderson-Holstein Hamiltonian, which reads
\begin{eqnarray}
H=\sum_\sigma\varepsilon_\sigma \hat{d}^\dagger_\sigma \hat{d}_\sigma+U_c\hat{d}^\dagger_\uparrow \hat{d}_\uparrow \hat{d}^\dagger_\downarrow \hat{d}_\downarrow+\varepsilon_p\hat{b}^\dagger \hat{b}+M\hat{n}(\hat{b}^\dagger+\hat{b})\nonumber\\
+\sum_{k,\sigma,\alpha}\varepsilon_k\hat{C}^\dagger_{k\sigma\alpha}\hat{C}_{k\sigma\alpha}+\sum_{k,\sigma,\alpha}\big(V_\alpha \hat{C}^\dagger_{k\sigma\alpha}\hat{d}_\sigma+\textrm{H.c.}\big),
\end{eqnarray}
where $\hat{d}^\dagger_\sigma$ ($\hat{C}^\dagger_{k\sigma\alpha}$) creates an electron with spin $\sigma=\uparrow,\downarrow$ and energy $\varepsilon_\sigma$ ($\varepsilon_k$) in the QD orbital (in the $\alpha$ lead, $\alpha=L,R$), and $U_c$ denotes the on-site Coulomb repulsion. Mechanical vibrations with frequency $\varepsilon_p$ are excited by the phonon operator $\hat{b}^\dagger$, which couple to the total dot charge $\hat{n}=\sum_\sigma\hat{d}^\dagger_\sigma\hat{d}_\sigma$ through a Holstein coupling strength $M$. Finally, electronic tunneling between the dot and leads is accounted for by tunneling matrix elements $V_\alpha$ in the last term. We then apply a polaronic transformation \cite{Mahan2000} $H\rightarrow\widetilde{H}=e^SHe^{-S}$, with $S=(M/\varepsilon_p)\hat{n}(\hat{b}^\dagger-\hat{b})$, to eliminate the electron-phonon coupling. As a result, the orbital energy gets renormalized $\varepsilon_\sigma\rightarrow\varepsilon_\sigma-M^2/\varepsilon_p$, which is canceled by redefining $\varepsilon_\sigma$. Due to the polaronic shift $U_c\rightarrow U=U_c-2M^2/\varepsilon_p$, the Coulomb repulsion can be renormalized downward to a bipolaronic attraction, realizing the scenario of negative charging energy. The transformation also leads to dressed tunneling matrix elements $V_\alpha\rightarrow V_\alpha e^{(M/\varepsilon_p)(\hat{b}-\hat{b}^\dagger)}$. For temperatures and biases lower than the phonon frequency, vibrational excitations are energetically not allowed. In this regime, after averaging over the zero-phonon state and redefining the dressed tunneling matrix elements as $V_\alpha$, the effective Hamiltonian $\widetilde{H}$ becomes the standard Anderson form
\begin{eqnarray}
\widetilde{H}&=&\sum_\sigma\varepsilon_\sigma \hat{d}^\dagger_\sigma \hat{d}_\sigma+U\hat{d}^\dagger_\uparrow \hat{d}_\uparrow \hat{d}^\dagger_\downarrow \hat{d}_\downarrow+\sum_{k,\sigma,\alpha}\varepsilon_k\hat{C}^\dagger_{k\sigma\alpha}\hat{C}_{k\sigma\alpha}\nonumber\\
&&+\sum_{k,\sigma,\alpha}\big(V_\alpha \hat{C}^\dagger_{k\sigma\alpha}\hat{d}_\sigma+\textrm{H.c.}\big),
\end{eqnarray}
but with the negative bipolaronic interaction $U$. Below we address the transport properties of QDs described by this Hamiltonian.

Within the Keldysh formalism \cite{Haug2008}, the electronic current through our system in the wide-band limit is $I=\frac{4e}{h}\frac{\Gamma_L\Gamma_R}{\Gamma}\int\textrm{d}\varepsilon\,[f_R(\varepsilon)-f_L(\varepsilon)]\sum_\sigma\textrm{Im}G^r_\sigma(\varepsilon)$, which expresses the current as an integral of the elastic transmission probability over the conduction band weighted by the difference of the Fermi functions $f_\alpha(\varepsilon)$ in the two leads. The transmission probability is constructed in terms of the dot retarded Green function  $G^r_\sigma(\varepsilon)\equiv\langle\langle\hat{d}^\dagger_\sigma;\hat{d}_\sigma\rangle\rangle$, and the dot level broadening due to tunnel coupling to the leads, $\Gamma=\sum_\alpha\Gamma_\alpha$ with $\Gamma_\alpha=\pi\sum_k|V_\alpha|^2\delta(\mu_\alpha-\varepsilon_k)$ calculated at the lead Fermi energy $\mu_\alpha$.

To determine the Green function $G^r_\sigma(\varepsilon)$, we employ
the equation-of-motion (EOM) approach
\cite{Meir1991,Haug2008,Groshev1991,Wohlman2005,sun1,sun2}. The EOM
for $G^r_\sigma(\varepsilon)$ gives rise to higher-order Green
functions, whose EOMs generate in turn more higher-order ones.
Provided that proper decoupling procedures have been used to
truncate this hierarchy, the approach can work in all parameter
regimes. For our purpose, we neglect those higher-order Green
functions which involve the spin-exchange scattering and
simultaneous creation or annihilation of two electrons in the QD
orbital, thereby excluding the contributions from the Kondo effect,
pair tunneling and cotunneling \cite{sun2}. By this scheme, we derive
the dot Green function as
\begin{equation}
G^r_\sigma(\varepsilon)=\frac{1-n_{\bar\sigma}}{\varepsilon-\varepsilon_\sigma+i\Gamma}+\frac{n_{\bar\sigma}}{\varepsilon-\varepsilon_\sigma-U+i\Gamma},
\end{equation}
where the occupation number $n_{\sigma}\equiv\langle\hat{d}^\dagger_\sigma\hat{d}_\sigma\rangle$ should be calculated from the nonequilibrium lesser Green function for the dot, $n_\sigma=-(i/2\pi)\int\textrm{d}\varepsilon\,G^<_\sigma(\varepsilon)$. Applying the formal Keldysh Green function technique \cite{Haug2008} to our system, the lesser Green function is related to the retarded one through $G^<_\sigma(\varepsilon)=-(2i/\Gamma)\sum_\alpha\Gamma_\alpha f_\alpha(\varepsilon)\textrm{Im}G^r_\sigma(\varepsilon)$. We can thus self-consistently calculate $n_\sigma$ and $G^r_\sigma(\varepsilon)$, from which the current $I$ or differential conductance $\textrm{d}I/\textrm{d}V$ can be obtained. Equation (3) shows that in the SE resonant tunneling regime, the QD has two resonances with width $\Gamma$: one at $\varepsilon_\sigma$ weighted by $1-n_{\bar\sigma}$, and one at $\varepsilon_\sigma+U$ weighted by $n_{\bar\sigma}$. A similar solution for $G^r_\sigma(\varepsilon)$ at positive $U$ was previously derived for explaining the periodic conductance oscillations \cite{Meir1991} and charging effects \cite{Groshev1991} in QDs. Here, we employ Eq.\,(3) to demonstrate the bipolaronic blockade effect due to the bipolaronic attraction between electrons, as a counterpart to the Coulomb blockade physics from the Coulomb repulsion. Despite its simplicity, the involved transport phenomena are unexpected and even surprising.

\section{III. results and discussions}
In the numerical results presented below, the half bandwidth $D=10$ is taken as the energy unit. We apply a symmetric source-drain voltage $V$ on the two leads whose Fermi energies are $\mu_\alpha=\mu+(\delta_{\alpha L}-\delta_{\alpha R})V/2$ with $\mu=0$ the equilibrium Fermi energy. The dot level $\varepsilon_\sigma$ is written as $\varepsilon_\sigma=\varepsilon_d+(\delta_{\sigma\uparrow}-\delta_{\sigma\downarrow})B$ in which $\varepsilon_d$ is the part tuned by the gate voltage and $B$ is the applied magnetic field. We introduce $x\equiv\Gamma_L/\Gamma_R$ to measure the left-right asymmetry of coupling and $y\equiv4x/(1+x)^2$ as a dimensionless measure of conductance in the asymmetric case. The temperature $T=10^{-8}D$ is set as the smallest energy scale in this work, though it should be higher than the underlying Kondo scale.

We first give, in Fig.\,1, the dot spectral density $A(\varepsilon)=-\frac{1}{\pi}\sum_\sigma\textrm{Im}G^r_\sigma(\varepsilon)$ and the occupation $n=\sum_\sigma n_\sigma$ in the equilibrium case for vanishing magnetic field. As the dot level $\varepsilon_d$ is tuned downward across the Fermi energy, the spectral weight of the $U<0$ dot shifts to the low-energy side of the Fermi energy more rapidly than that of $U>0$ [Figs.\,1(a)-1(c)], thereby leading to a rapid increase of the occupation [Fig.\,1(d)]. Particularly, the occupation governed by the bipolaronic attraction increases linearly with $\varepsilon_d$ near the particle-hole symmetric point, where the Coulomb repulsion induces the usual plateau. To give a quantitative account for this difference, we calculate the occupation $n$ to first order in the deviation from the particle-hole symmetric point $\delta\varepsilon_d=\varepsilon_d+U/2$, which is, in the limit $T/\Gamma\rightarrow0$, $n=1-8\Gamma(U^2+4\Gamma^2)^{-1}[\pi+2\arctan(\frac{U}{2\Gamma})]^{-1}\delta\varepsilon_d$. For $\Gamma\ll|U|$, the attraction results in $n=1-\frac{2}{|U|}\delta\varepsilon_d$ and the repulsion gives $n=1$. Fig.\,1(d) also indicates that the bipolaronic attraction significantly suppresses the dot spectral density at the Fermi energy as compared with the Coulomb repulsion. These unusual local properties of the $U<0$ dot can bring about very striking transport behavior when the source-drain voltage is applied.

\begin{figure}
\centering
\includegraphics[width=1.0\columnwidth]{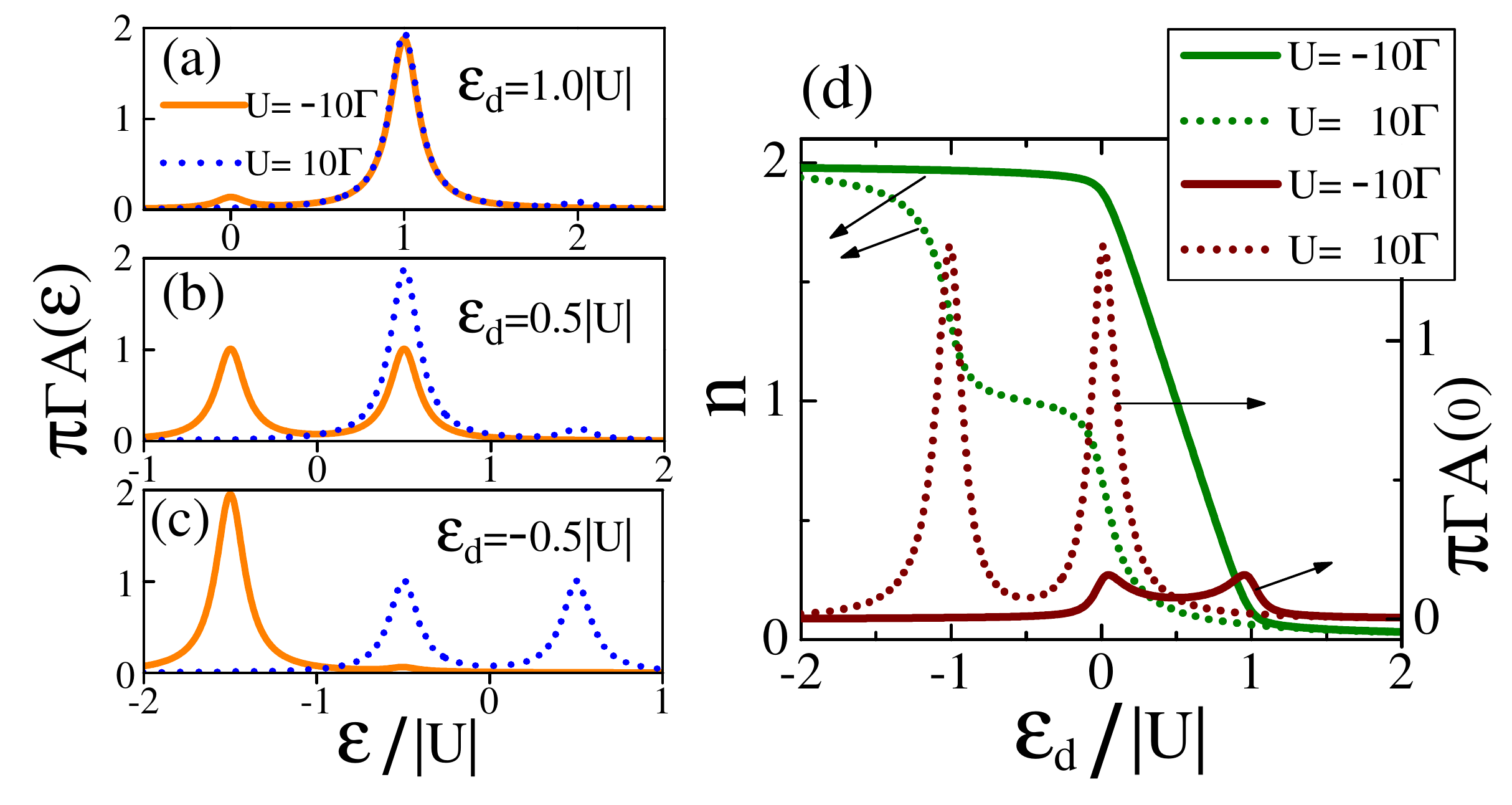}
\caption{(a)-(c) Local spectral densities $A(\varepsilon)$ as a function of energy $\varepsilon$ for different positions of the dot level $\varepsilon_d$. (d) Dot occupation number $n$ and spectral density at the Fermi energy $A(0)$ as a function of the dot level $\varepsilon_d$. Results for positive $U$ are also presented for comparison. Parameters used are $V=0$, $\Gamma=10^{-3}D$, and $B=0$.}
\end{figure}

Figure 2(a) presents the conductance map subject to the bipolaronic attraction for symmetric couplings and zero magnetic field, where the conductance lines appear at the thresholds $V$ and $\varepsilon_d$ that switch the system between the configurations of two (L-$2$), one (L-$1$), and zero (L-$0$) QD level(s) in the transport window. A striking feature distinct from the usual Coulomb-blockade spectroscopy \cite{Kastner1992,Yang2005,Bockrath1997,Zimbovskaya2011,Meir1991} is that the conductance lines under low biases $|V|<|U|$ is now suppressed. This bipolaronic blockade effect becomes more prominent for weak dot-lead coupling $\Gamma$ [Figs.\,2(b)-2(f)]. Particularly, in the limit $\Gamma\ll|U|$ the linear conductance is completely suppressed [Fig.\,2(b)], eliminating the usual periodic Coulomb oscillations at $\varepsilon_d=0$ and $|U|$. Note that the conductance lines at $|V|<|U|$ represent the transitions between the L-$0$ and L-$1$ configurations. While SE tunneling is obviously not allowed in the L-$0$ configuration since there has no QD levels in the transport window, the tunneling is also suppressed in the L-$1$ configuration where the negative $U$ favors an empty [Fig.\,2(g)] or doubly-occupied [Fig.\,2(h)] QD with very small spectral weight left in the transport window. Therefore, no significant current change occurs between the two configurations, leading to the suppressed conductance at low biases. On the other hand, the L-$2$ configuration always allows two independent SE tunnelings [Fig.\,2(i)] giving the maximal current $I_0\equiv ye\Gamma/\hbar$. As a result, the conductance lines for large biases $|V|>|U|$, representing the transitions between the L-$1$ and L-$2$ configurations, are thus enhanced.

\begin{figure}
\centering
\includegraphics[width=1.0\columnwidth]{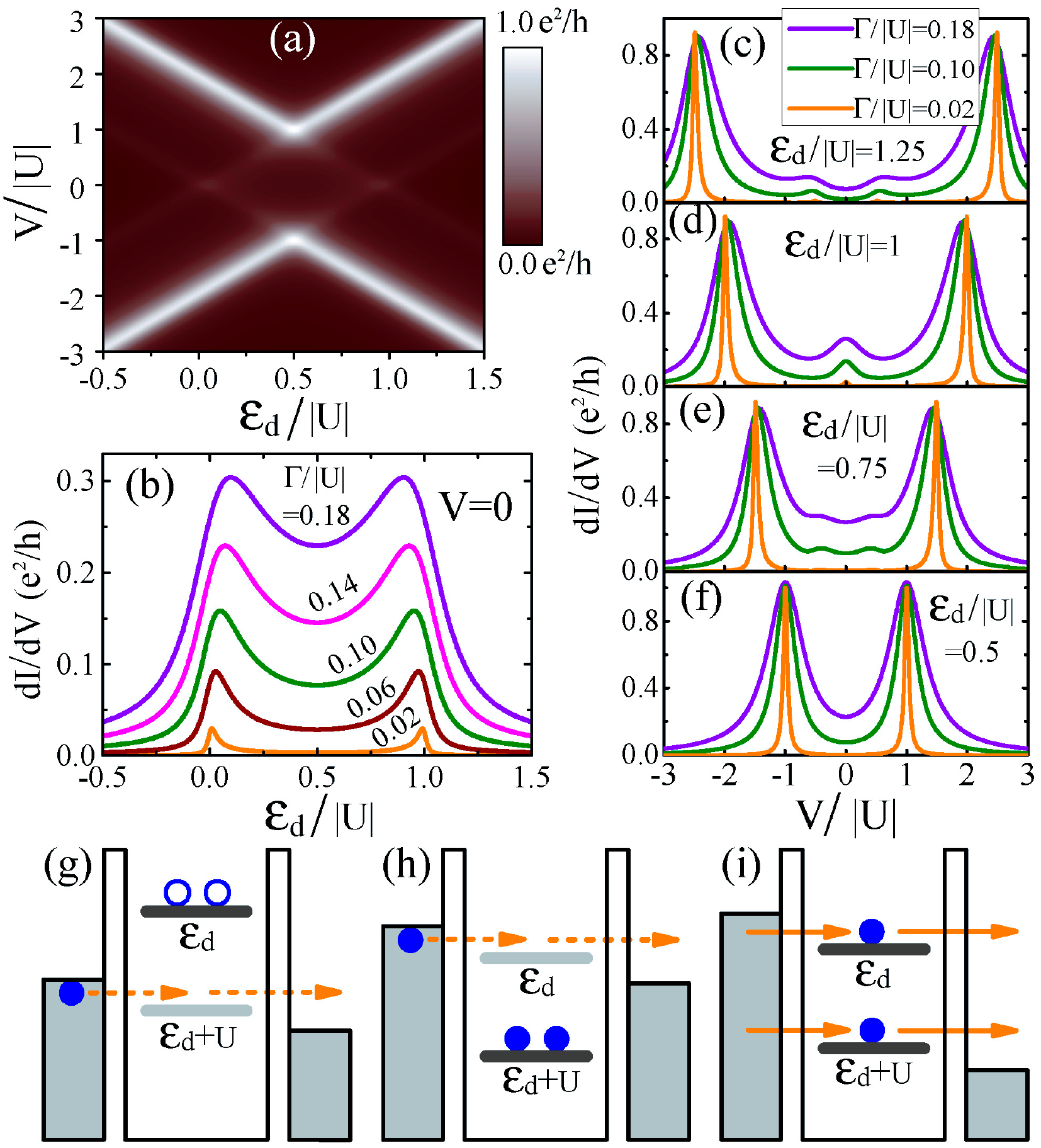}
\caption{(a) Differential conductance map as a function of the dot level $\varepsilon_d$ and source-drain bias $V$ for symmetric coupling $\Gamma=0.1|U|$. (b) Linear conductance for different $\Gamma$. (c)-(f) Conductance cut at several $\varepsilon_d$ for different $\Gamma$. (g)-(i) Schematic diagrams of one dot level in the transport window suppressing SE tunnelings [(g) and (h)] and two dot levels in the transport window allowing the tunneling (i), where the solid and hollow circles represent electrons and holes, respectively. Parameters used are $U=-0.1$, $x=1$, and $B=0$.}
\end{figure}

When asymmetric coupling is introduced, as shown in Fig.\,3(a) for $x>1$, the conductance lines enclosing the L-$1$ regions with positive (or negative if $x<1$) slope merge into finite plateaus. The plateau becomes concrete for increasing the asymmetry [Fig.\,3(b)] and is elevated by strengthening the dot-lead coupling [Fig.\,3(c)]. Its height in the middle is estimated to be $(\pi\Gamma/|U|)ye^2/h$ in the asymmetric limit. To explain the formation of the plateau, let us consider a given electronic state $\varepsilon_d=|U|$. As the source-drain bias rises positively, the occupation number $n$ increases linearly for $x\gg1$ [Fig.\,3(d)], since in this limit $n$ is only determined by the left lead and is equivalent to the equilibrium one [Fig.\,1(d)]. Accordingly, the QD spectral weight shifts linearly from the level $\varepsilon_d$ to the level $\varepsilon_d+U$ which lies in the transport windows [Fig.\,3(e)] giving rise to a linear increase of the current. Therefore, the conductance is constant until the Fermi level of the left lead $\mu_L$ sweeps over the level $\varepsilon_d$ saturating the current. On the other hand, as the bias rises negatively, the QD spectral density in the transport window and hence the current keep to be nearly zero until $\mu_R$ sweeps over the level $\varepsilon_d$ [Fig.\,3(f)], which leads to the conductance peak at $V=-2|U|$ [see Figs.\,3(a)-3(c)].

\begin{figure}
\centering
\includegraphics[width=1.0\columnwidth]{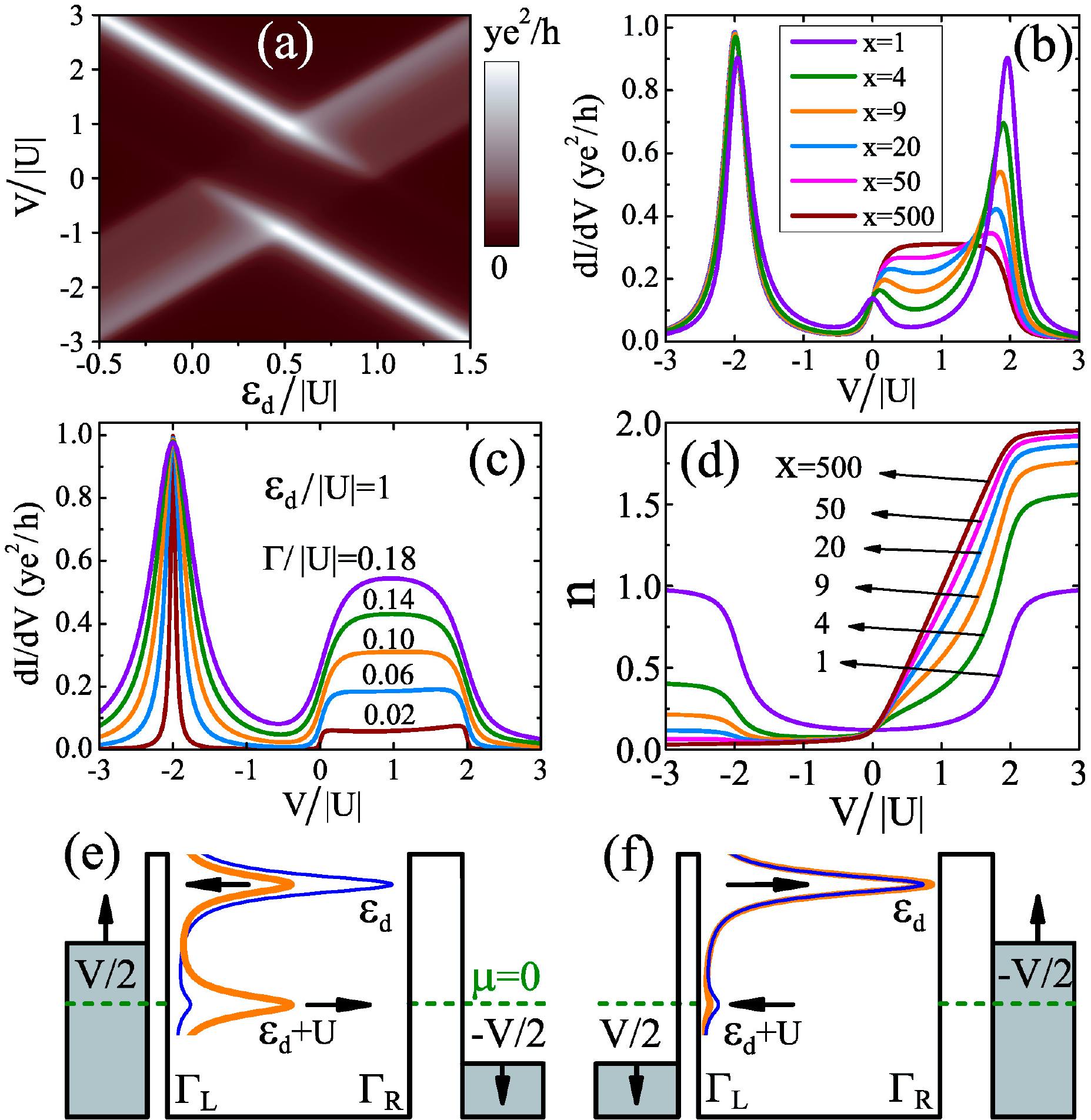}
\caption{(a) Conductance map for asymmetric coupling $\Gamma=0.1|U|$, $x=20$. (b) and (c) Conductance vs $V$ for different $x$ with $\Gamma=0.1|U|$ and different $\Gamma$ with $x=500$, respectively. (d) Occupation number vs $V$ for different $x$ with $\Gamma=0.1|U|$. (e) and (f) Schematic diagrams for the shift of QD spectral weight in the asymmetric limit when $V$ increases positively (e) and negatively (f), where the blue (orange) lines corresponds to the spectral densities at equilibrium (at indicated $V$). Parameters: $\varepsilon_d=|U|$ for (b)-(f), $U=-0.1$, $B=0$.}
\end{figure}

We now turn to examine the effect of magnetic field, which splits
the conductance peaks and narrows the plateaus [Figs.\,4(a)-4(c)].
While the narrowing of the plateaus is quite straightforward by just
noting that at fixed $\varepsilon_d$ and $x\gg1$ the plateau starts
and ends when $\mu_L$ crosses the levels $\varepsilon_\uparrow+U$
and $\varepsilon_\downarrow$, the feature of split conductance peaks
deserves a careful analysis. For symmetric couplings, the amplitudes
of split peaks are different, reaching a ratio of $3$ in the weak
coupling limit [Fig.\,4(d)]. This is different from the Coulomb
Blockade physics where the corresponding ratio is
$2$ \cite{Cobden1998}. The underlying mechanism can be revealed by
analyzing the energy-level diagrams of Fig.\,5 at
$\varepsilon_d=|U|$ and characteristic biases marked by squares,
circles, and triangles in Figs\,4(a) and 4(b). Note that at finite
magnetic field the spectral weight of the four QD levels
$\varepsilon_\uparrow$, $\varepsilon_\downarrow$,
$\varepsilon_\uparrow+U$, and $\varepsilon_\downarrow+U$ are
proportional to the occupation $1-n_\downarrow$, $1-n_\uparrow$,
$n_\downarrow$, and $n_\uparrow$, respectively, as shown by
Eq.\,(3). Since $n_\sigma\simeq0$ for the negative bias at the
square point [Fig.\,5(a)], no significant spectral density is
enclosed in the transport window giving $I\simeq0$. As the bias
further decreases to the circle point [Fig.\,5(b)], one has
$n_\uparrow\simeq0.25$ and $n_\downarrow\simeq0.5$. This produces a
current $I=0.75I_0$, in which the portion $0.5I_0$ is the
contribution from the two spin-down levels $\varepsilon_\downarrow$
and $\varepsilon_\downarrow+U$, and $0.25I_0$ from the spin-up level
$\varepsilon_\uparrow+U$. The current finally saturates to $I_0$
under the bias at the triangle point where the total QD spectral
density contributes [Fig.\,5(c)]. Therefore, two conductance peaks
with the amplitude ratio of $3$ appear as $\mu_R$ successively
sweeps over the Zeeman-split levels $\varepsilon_\downarrow$ and
$\varepsilon_\uparrow$. Similar analysis can indicate that the ratio
turns to $1$ in the asymmetric limit [Fig.\,4(d)]. In this case, the
occupation is only determined by the stronger coupled lead and one
always have $n_\sigma\simeq0$, which results in the current
$I\simeq0$ [Fig.\,5(d)], $0.5I_0$ [Fig.\,5(e)], and $I_0$
[Fig.\,5(f)] at the three characteristic biases, respectively.

\begin{figure}
\centering
\includegraphics[width=1.0\columnwidth]{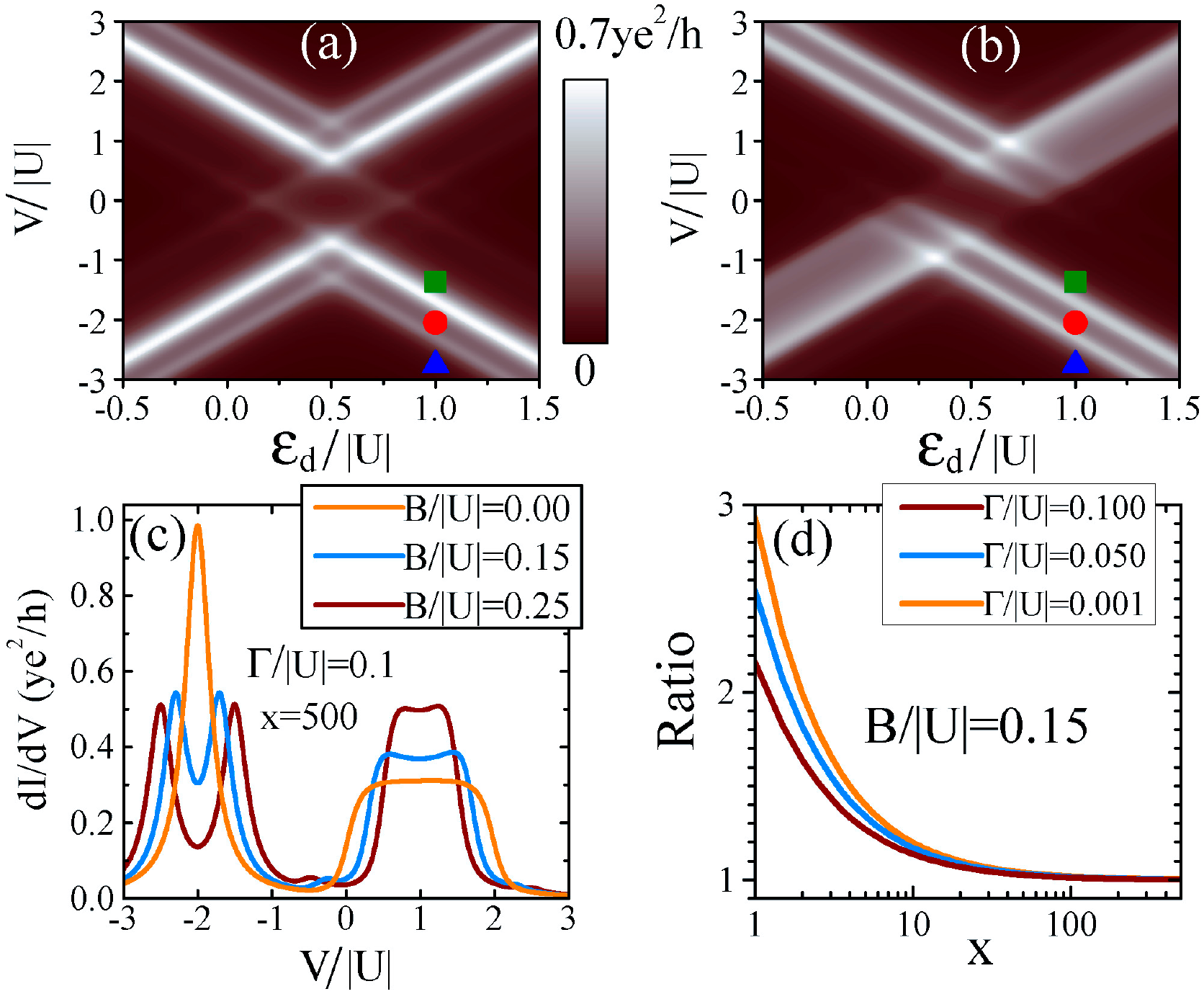}
\caption{Conductance map under magnetic field $B=0.15|U|$ for (a) symmetric coupling $x=1$ and (b) asymmetric coupling $x=20$. (c) Lines cut from (b) at $\varepsilon_d=|U|$ for different magnetic fields and $x=500$. (d) Amplitude ratio of magnetic-split conductance peaks as a function of the coupling asymmetry. In (a) and (b), the squares, circles, and triangles denote three representative biases at $\varepsilon_d=|U|$. Other parameters: $\Gamma=0.1|U|$ for (a)-(c) and $U=-0.1$.}
\end{figure}

\begin{figure}
\centering
\includegraphics[width=1.0\columnwidth]{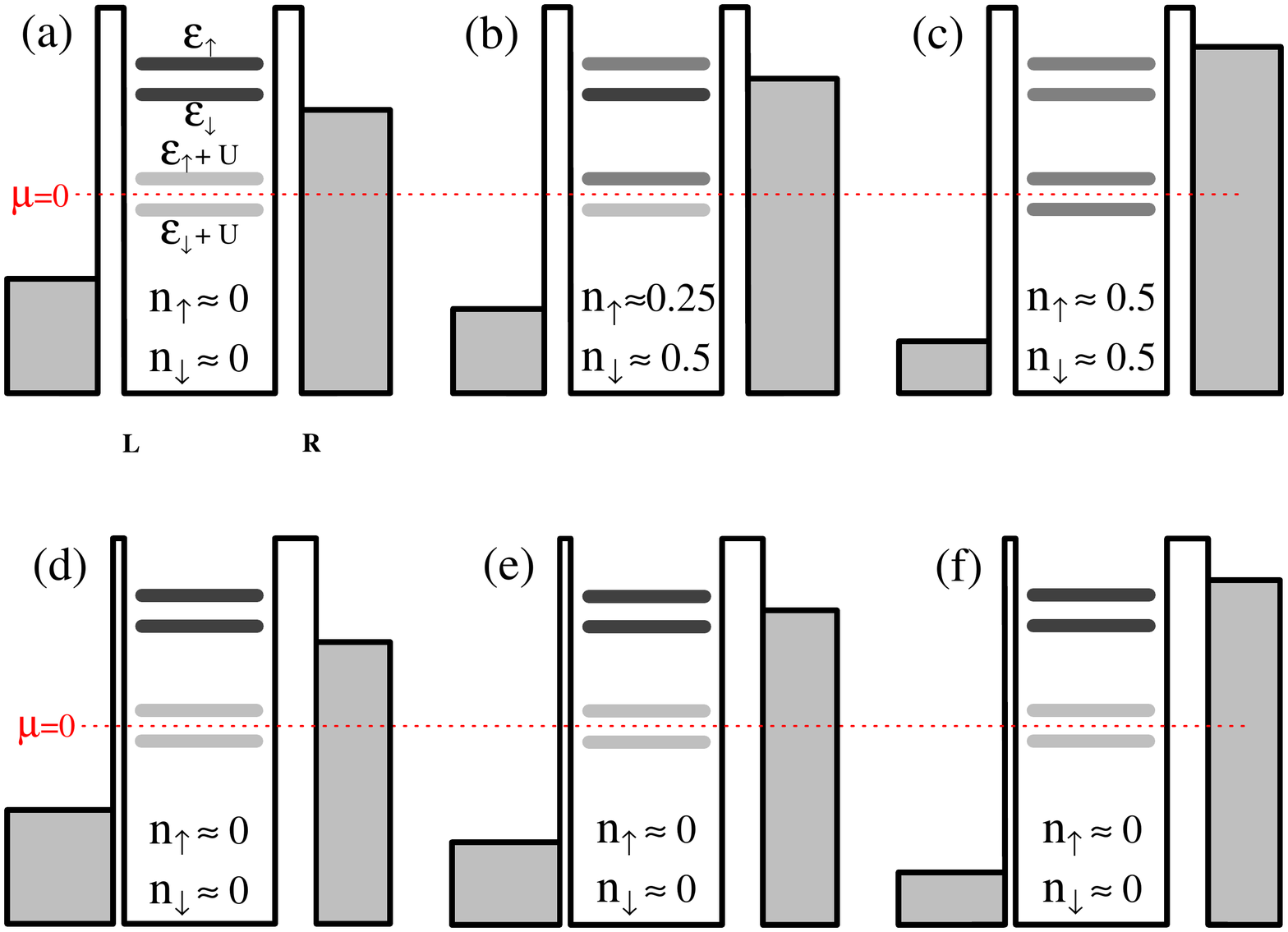}
\caption{Energy-level diagrams for symmetric $x=1$ [(a)-(c)] and asymmetric $x\gg1$ [(d)-(f)] couplings with finite magnetic field, under the source-drain biases corresponding to the squares [(a) and (d)], circles [(b) and (e)], and triangles [(c) and (f)] in Fig.\,4. The light gray, gray, and dark gray lines represent the four QD levels with small, moderate, and large spectral weights, respectively.}
\end{figure}

\section{IV. final remarks}
Finally, it is helpful to comment on the observability of these transport phenomena for negative charging energy. A conceivable realization might be to use suspended carbon-nanotube devices, for which strong electron-phonon coupling has already been observed \cite{SCNT}. The most favorable candidate to cause a negative $U$ in nanotubes is the longitudinal stretching mode with the maximal dimensionless coupling $\lambda\equiv\frac{M}{\varepsilon_p}\sim3/\sqrt{L_\perp[\textrm{nm}]}$ \cite{Mariani2009} and the phonon energy $\varepsilon_p\sim\frac{0.11\textrm{meV}}{L[\mu\textrm{m}]}$ \cite{Sapmaz2006}, where $L_\perp$ is the nanotube circumference and $L$ the length. Using $L_\perp\sim1\textrm{nm}$ and $L\sim100\textrm{nm}$, we estimate the polaronic energy shift as $2\lambda^2\varepsilon_p\sim20\textrm{meV}$, being of the same order as or larger than typical Coulomb repulsion in nanotube QDs \cite{Kuemmeth2008}. After realizing $U<0$, transport spectroscopy measurements similar to the usual Coulomb-blockade technique are enough to verify our predictions, including the suppression of conductance at low biases, the appearance of conductance plateaus for asymmetric coupling, and the ratio of $3$ for magnetic-split conductance peaks.

\section{acknowledgments}
This work was financially supported by NBRP of China (2012CB921303)
and NSF-China under Grant Nos. 11004090 and 11274364.


\begin{thebibliography}{10}
\bibitem{Kastner1992}
M.\,A. Kastner, Rev. Mod. Phys. \textbf{64}, 849 (1992); L.\,P. Kouwenhoven and C.\,M. Marcus, Phys. World \textbf{11}, 35 (1998); I.\,L. Aleiner, P.\,W. Brouwer, L.\,I. Glazman, Phys. Rep. \textbf{358}, 309 (2002); S.\,M. Reimann and M. Manninen, Rev. Mod. Phys. \textbf{74}, 1283 (2002).
\bibitem{Yang2005}
C. Yang, Z. Zhong, and C.\,M. Lieber, Science \textbf{310}, 1304 (2005); S. Nadj-Perge, S.\,M. Frolov, E.\,P.\,A.\,M. Bakkers, and L.\,P. Kouwenhoven, Nature (London) \textbf{468}, 1084 (2010).
\bibitem{Bockrath1997}
M. Bockrath, D.\,H. Cobden, P.\,L. McEuen, N.\,G. Chopra, A. Zettl, A. Thess, and R.\,E. Smalley, Science \textbf{275}, 1922 (1997); J. Nyg{\aa}rd, D.\,H. Cobden, and P.\,E. Lindelof, Nature (London) \textbf{408}, 342 (2000).
\bibitem{Zimbovskaya2011}
N.\,A. Zimbovskaya and M.\,R. Pederson, Phys. Rep. \textbf{509}, 1 (2011).
\bibitem{Meir1991}
Y. Meir, N.\,S. Wingreen, and P.\,A. Lee, Phys. Rev. Lett. \textbf{66}, 3048 (1991).
\bibitem{SM}
H. Park, J. Park, A.\,K.\,L. Lim, E.\,H. Anderson, A.\,P. Alivisatos, and P.\,L. McEuen, Nature (London) \textbf{407}, 57 (2000); J. Park, A.\,N. Pasupathy, J.\,I. Goldsmith, C. Chang, Y. Yaish, J.\,R. Petta, M. Rinkoski, J.\,P. Sethna, H.\,D. Abru\~{n}a, P.\,L. McEuen, and D.\,C. Ralph, Nature (London) \textbf{417}, 722 (2002); W. Liang, M.\,P. Shores, M. Bockrath, J.\,R. Long, and H. Park, Nature (London) \textbf{417}, 725 (2002); N.\,B. Zhitenev, H. Meng, and Z. Bao, Phys. Rev. Lett. \textbf{88}, 226801 (2002); L.\,H. Yu, Z.\,K. Keane, J.\,W. Ciszek, L. Cheng, M.\,P. Stewart, J.\,M. Tour, and D. Natelson, Phys. Rev. Lett. \textbf{93}, 266802 (2004); X.\,H. Qiu, G.\,V. Nazin, and W. Ho, Phys. Rev. Lett. \textbf{92}, 206102 (2004); J.\,J. Parks, A.\,R. Champagne, G.\,R. Hutchison, S. Flores-Torres, H.\,D. Abru\~{n}a, and D.\,C. Ralph, Phys. Rev. Lett. \textbf{99}, 026601 (2007); I. Fern\'{a}ndez-Torrente, K.\,J. Franke, and J.\,I. Pascual, Phys. Rev. Lett. \textbf{101}, 217203 (2008).
\bibitem{SCNT}
B.\,J. LeRoy, S.\,G. Lemay, J. Kong, and C. Dekker, Nature (London) \textbf{432}, 371 (2004); B.\,J. LeRoy, J. Kong, V.\,K. Pahilwani, C. Dekker, and S.\,G. Lemay, Phys. Rev. B \textbf{72}, 075413 (2005); R. Leturcq, C. Stampfer, K. Inderbitzin, L. Durrer, C. Hierold, E. Mariant, M.\,G. Schultz, F. von Oppen, and K. Ensslin, Nat. Phys. \textbf{5}, 327 (2009); A.\,K. H\"{u}ttel, B. Witkamp, M. Leijnse, M.\,R. Wegewijs, and H.\,S.\,J. van der Zant, Phys. Rev. Lett. \textbf{102}, 225501 (2009); F. Cavaliere, E. Mariani, R. Leturcq, C. Stampfer, and M. Sassetti, Phys. Rev. B \textbf{81}, 201303(R) (2010).
\bibitem{CK}
P.\,S. Cornaglia, H. Ness, and D.\,R. Grempel, Phys. Rev. Lett. \textbf{93}, 147201 (2004); P.\,S. Cornaglia, D.\,R. Grempel, and H. Ness, Phys. Rev. B \textbf{71}, 075320 (2005); L. Arrachea and M.\,J. Rozenberg, Phys. Rev. B \textbf{72}, 041301(R) (2005); J. Mravlje, A. Ram\v{s}ak, and T. Rejec, Phys. Rev. B \textbf{72}, 121403(R) (2005); J. Koch, E. Sela, Y. Oreg, and F. von Oppen, Phys. Rev. B \textbf{75}, 195402 (2007); J.\,S. Lim, R. L\'{o}pez, G. Platero, and P. Simon, Phys. Rev. B \textbf{81}, 165107 (2010).
\bibitem{Koch2006}
J. Koch, M.\,E. Raikh, and F. von Oppen, Phys. Rev. Lett. \textbf{96}, 056803 (2006).
\bibitem{Hwang2007}
M.-J. Hwang, M.-S. Choi, and R. L\'{o}pez, Phys. Rev. B \textbf{76}, 165312 (2007).
\bibitem{Wysokinski2010}
K.\,I. Wysoki\'{n}ski, Phys. Rev. B \textbf{82}, 115423 (2010).
\bibitem{SO}
Taking account of only one orbital is valid when the single-particle level spacing exceeds the magnitude of the negative charging energy and thus the electron tunneling is stable. In the opposite regime, the system becomes unstable toward tunneling of an arbitrary number of electrons into or out of the QD, as shown by T. Ojanen, F.\,C. Gethmann, and F. von Oppen in Phys. Rev. B \textbf{80}, 195103 (2009).
\bibitem{Mahan2000}
G.\,D. Mahan, \textsl{Many-Particle Physics}, 3rd ed.\,(Plenum, New York, 2000).
\bibitem{Haug2008}
H. Haug and A.-P. Jauho, \textsl{Quantum Kinetics in Transport and Optics of Semiconductors}, 2nd ed.\,(Springer, Berlin, 2008).
\bibitem{Groshev1991}
A. Groshev, T. Ivanov, and V. Valtchinov, Phys. Rev. Lett. \textbf{66}, 1082 (1991).
\bibitem{Wohlman2005}
O. Entin-Wohlman, A. Aharony, and Y. Meir, Phys. Rev. B \textbf{71},
035333 (2005); V. Kashcheyevs, A. Aharony, and O. Entin-Wohlman,
Phys. Rev. B \textbf{73}, 125338 (2006); Y. Meir, N.\,S. Wingreen,
and P.\,A. Lee, Phys. Rev. Lett. \textbf{70}, 2601 (1993).

\bibitem{sun1}
Q.-F. Sun, H. Guo, and T.-H. Lin, Phys. Rev. Lett. \textbf{87},
176601 (2001); Q.-F. Sun and H. Guo, Phys. Rev. B \textbf{66},
155308 (2002); N. Sergueev, Q.-F. Sun, H. Guo, B. G. Wang, and J.
Wang, Phys. Rev. B \textbf{65}, 165303 (2002).

\bibitem{sun2}
T.-F. Fang, Q.-F. Sun, and H.-G. Luo, Phys. Rev. B \textbf{84}, 155417 (2011); J. Liu, Q.-F. Sun, and X.-C. Xie, Phys. Rev. B \textbf{79}, 161309(R) (2009).

\bibitem{Cobden1998}
D.\,H. Cobden, M. Bockrath, and P.\,L. McEuen, Phys. Rev. Lett. \textbf{81}, 681 (1998); H. Akera, Phys. Rev. B \textbf{60}, 10683 (1999); J. Park, A.\,N. Pasupathy, J.\,I. Goldsmith, C. Chang, Y. Yaish, J.\,R. Petta, M. Rinkoski, J.\,P. Sethna, H.\,D. Abru\~{n}a, P.\,L. McEuen, and D.\,C. Ralph, Nature (London) \textbf{417}, 722 (2002); M.\,M. Deshmukh, E. Bonet, A.\,N. Pasupathy, and D.\,C. Ralph, Phys. Rev. B \textbf{65}, 073301 (2002); E. Bonet, M.\,M. Deshmukh, and D.\,C. Ralph, Phys. Rev. B \textbf{65}, 045317 (2002); S. Moriyama, T. Fuse, M. Suzuki, Y. Aoyagi, and K. Ishibashi, Phys. Rev. Lett. \textbf{94}, 186806 (2005).
\bibitem{Mariani2009}
E. Mariani and F. von Oppen, Phys. Rev. B \textbf{80}, 155411 (2009).
\bibitem{Sapmaz2006}
S. Sapmaz, P. Jarillo-Herrero, Ya.\,M. Blanter, C. Dekker, and H.\,S.\,J. van der Zant, Phys. Rev. Lett. \textbf{96}, 026801 (2006).
\bibitem{Kuemmeth2008}
J. Nyg{\aa}rd, D.\,H. Cobden, and P.\,E. Lindelof, Nature (London) \textbf{408}, 342 (2000);P. Jarillo-Herrero, J. Kong, H.\,S.\,J. van der Zant, C. Dekker, L.\,P. Kouwenhoven, and S.\,D. Franceschi, Nature (London) \textbf{434}, 484 (2005); F. Kuemmeth, S. Ilani, D.\,C. Ralph, and P.\,L. McEuen, Nature (London) \textbf{452}, 448 (2008); T.\,S. Jespersen, K. Grove-Rasmussen, J. Paaske, K. Muraki, T. Fujisawa, J. Nyg{\aa}rd, and K. Flensberg, Nat. Physics \textbf{7}, 348 (2011).
\end{thebibliography}
\end{document}